\documentclass[acmsmall,screen]{acmart}
\usepackage{enumitem}

\AtBeginDocument{%
  \providecommand\BibTeX{{%
    \normalfont B\kern-0.5em{\scshape i\kern-0.25em b}\kern-0.8em\TeX}}}

\setcopyright{acmlicensed}
\acmJournal{PACMHCI}
\acmYear{2020} \acmVolume{4} \acmNumber{CSCW1} \acmArticle{60} \acmMonth{5} \acmPrice{15.00}\acmDOI{10.1145/3392869}

\begin{document}
\newcommand{\tlrevision}[1]{{{#1}}} 
\title{Privacy-Preserving Script Sharing in GUI-based Programming-by-Demonstration Systems} 


\author{Toby Jia-Jun Li}
\affiliation{%
  \institution{Carnegie Mellon University}
  \streetaddress{}
  \city{Pittsburgh, PA}
  \country{USA}}
\email{tobyli@cs.cmu.edu}

\author{Jingya Chen}
\affiliation{%
  \institution{Carnegie Mellon University}
  \streetaddress{}
  \city{Pittsburgh, PA}
  \country{USA}}
\email{jingyach@andrew.cmu.edu}

\author{Brandon Canfield}
\authornote{Work done as a visiting research assistant at Carnegie Mellon University.}
\affiliation{%
  \institution{Yale University}
  \streetaddress{}
  \city{New Haven, CT}
  \country{USA}}
\email{brandon.canfield@yale.edu}

\author{Brad A. Myers}
\affiliation{%
  \institution{Carnegie Mellon University}
  \streetaddress{}
  \city{Pittsburgh, PA}
  \country{USA}}
\email{bam@cs.cmu.edu}


\begin{abstract}
An important concern in end user development (EUD) is accidentally embedding personal information in program artifacts when sharing them. This issue is particularly important in GUI-based programming-by-demonstration (PBD) systems due to the lack of direct developer control of script contents. Prior studies reported that these privacy concerns were the main barrier to script sharing in EUD. We present a new approach that can identify and obfuscate the potential personal information in GUI-based PBD scripts based on the \textit{uniqueness} of information entries with respect to the corresponding app GUI context. Compared with the prior approaches, ours supports broader types of personal information beyond explicitly pre-specified ones, requires minimal user effort, addresses the threat of re-identification attacks, and can work with third-party apps from any task domain. Our approach also recovers obfuscated fields locally on the script consumer's side to preserve the shared scripts' transparency, readability, robustness, and generalizability. Our evaluation shows that our approach (1) accurately identifies the potential personal information in scripts across different apps in diverse task domains; (2) allows end-user developers to feel comfortable sharing their own scripts; and (3) enables script consumers to understand the operation of shared scripts despite the obfuscated fields.
\end{abstract}



\maketitle

\section{Introduction}
\label{sec:intro}

The concept of end-user development (EUD), according to a popular definition, refers to programming activities to achieve results primarily intended for personal, rather than public use~\cite{ko_state_2011}. It allows individual users to develop and adapt systems according to their own needs and preferences~\cite{lieberman_end-user_2006}. Despite the ``for-self-use'' main intention, sharing of the program artifacts is common in EUD (e.g.,~\cite{Mackay:1990:PSC:99332.99356}). However, when sharing EUD artifacts, users are often concerned about privacy: they do not want to leak any personal information that may be embedded in the shared artifacts~\cite{leshed_coscripter:_2008,li_heres_2010}. 

\begin{figure}
	\centering
	\includegraphics[width=\columnwidth]{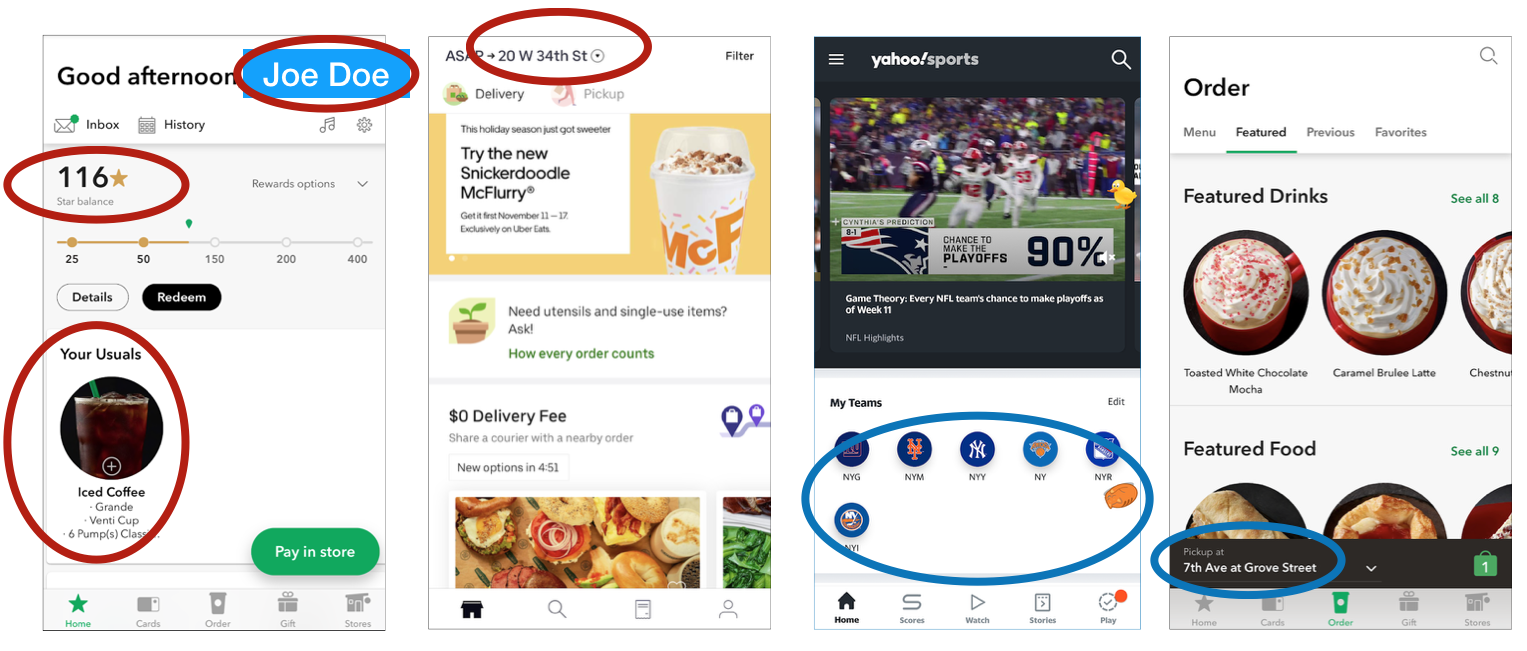}
	\caption{Some examples of potential information leaks from the app GUIs. The red circles highlight personal information directly displayed (e.g., name, point balance, order history, current address). The blue circles highlight information that can be used for re-identification attacks (e.g., the user's favorite teams and the name of a nearby Starbucks location can be used for inferring the user's location).}~\label{fig:example_leaks}
	\vspace{-0.5cm}
\end{figure}

This concern is particularly important in graphical user interface (GUI) based programming by demonstration (PBD) systems. In such systems, end-user developers do not directly author the programs, but instead demonstrate one or more examples of the desired system behaviors using the GUIs of target applications. From the demonstrations, the system synthesizes a program that can invoke and control the target applications to perform tasks. As a result, end-user developers have little knowledge or control about what has been included in the resulting program, and therefore are hesitant about sharing them~\cite{bogart_end-user_2008, leshed_coscripter:_2008, li_heres_2010}. In the recent GUI-based PBD agents for task automation (e.g., \textsc{Sugilite}~\cite{li_sugilite:_2017},  \textsc{Pumice}~\cite{li_pumice:_2019}, and \textsc{VASTA}~\cite{sereshkeh2020vasta}), the system not only collects information about the exact demonstrated actions, but also records rich contextual information (e.g., the contents of the screens including the app's responses) in order to better infer the users' intents from the demonstrations and to further generalize the resulting programs. While these new mechanisms make the PBD more powerful, flexible and robust, they also increase the likelihood of inappropriately including personal information in program artifacts, and make it infeasible for end-user developers to manually scrutinize the program artifacts before sharing them, since the collected contextual meta information is large in amount (averaging over 100 strings per script in our tests) and not generally shown to users. Many information fields displayed in app GUIs are dynamically generated based on the user's personal information. Figure~\ref{fig:example_leaks} shows some examples of personal information displayed in app GUIs that may be included in PBD scripts.

Another constraint in removing personal information from program artifacts created from GUI demonstrations is to preserve the scripts' transparency, readability, robustness, and generalizability. In task automation, it is important to enable users who plan to use a script shared by others to be able to fully understand its behavior, so that they can (1) validate if the script actually fulfills their needs, (2) identify any potentially dangerous or malicious operations in the script, and (3) modify or extend the script when needed. As a result, the system should avoid unnecessary obfuscation of the shared scripts that hinders transparency and readability. The system should also preserve as much meta information from demonstrations as possible in shared scripts, since such meta information is very useful for recovering from script execution errors, extending existing scripts, and supporting effective parameterization in scripts (see more details in Section~\ref{sec:gui_graph}).

There is also the emerging challenge of preventing \textit{re-identification attacks}~\cite{el2011systematic, benitez2010evaluating} when protecting identifiable personal information. In many cases, user identities and private information can be re-identified by combining multiple seemingly innocent data sources. For example, knowing seemingly non-private information like the user's local sports teams, the default Starbucks store, and the estimated wait time for food delivery, one can infer the user's location fairly accurately (Figure~\ref{fig:example_leaks} shows some examples). This is also known as a \textit{record linkage} attack~\cite{vatsalan2013taxonomy}. Prior approaches that mask only pre-specified types of explicit personal information (e.g.,~\cite{continella2017obfuscation, Wang:2019:LTA:3323054.3314415, leshed_coscripter:_2008}) are not sufficient when dealing with re-identification attacks.\looseness=-1

In this paper, we present a new mechanism named \textsc{Pinalite}\footnote{\textsc{Pinalite} is named after a type of rock. It is also an acronym for \textbf{P}ersonal \textbf{I}nformation \textbf{N}icely \textbf{A}nonymized \textbf{L}everaging \textbf{I}nterface \textbf{T}race \textbf{E}xamples.} for privacy-preserving sharing of GUI-based PBD task automation scripts. \textsc{Pinalite}'s approach can identify and obfuscate possible personal information at the time of sharing with minimal user intervention. This approach collects and aggregates hashed text strings from app GUIs on users' phones in everyday use. This aggregated data allows our system to identify fields in end-user-developed scripts with potentially personal information, and obfuscate their values by hashing. Unlike prior systems, \textsc{Pinalite} works with third-party apps from any task domain without prior knowledge about the app or the domain, and can process broad types of personal information beyond explicitly pre-specified ones to protect against re-identification attacks. At runtime, the obfuscated fields can be rebuilt locally using the script consumer's app GUIs, which helps preserve the shared scripts' transparency, readability, robustness, and generalizability. 

\section{Related Work}

\subsection{GUI-based Programming by Demonstration Systems}
\textsc{Pinalite} is applied to GUI-based programming by demonstration (PBD) task automation systems. These systems allow users to program automation scripts (also known as \textit{macros}) by demonstrating performing the target tasks using existing GUIs. This approach lowers the learning barriers~\cite{lieberman_your_2001, cypher_watch_1993} and improves naturalness~\cite{myers_natural_2004, myers_making_2017, li_teaching_2018} for end user developers because they do not have to directly author scripts using textual or visual programming languages. Instead, they can demonstrate the tasks with familiar existing third-party GUIs. However, a major limitation of PBD with respect to privacy is that users have little knowledge or control about what gets included in the scripts if they are shared, because the scripts are inferred from demonstrations by the system instead of authored directly by users.

Many GUI-based PBD systems have been developed for desktop (e.g., \cite{yeh_sikuli:_2009, intharah_help_2017}), mobile (e.g., \cite{li_sugilite:_2017}), and web interfaces (e.g., \cite{leshed_coscripter:_2008, li_heres_2010, chasins2017skip, chasins2018rousillon, allen_plow:_2007}). Among them, some are domain-general (e.g., ~\cite{li_sugilite:_2017, allen_plow:_2007, leshed_coscripter:_2008, Intharah:2019:HDP:3320251.3234508}), while others are designed for specific task domains such as data scraping~\cite{ chasins2017skip, chasins2018rousillon}, text editing~\cite{lau_programming_2003}, photo editing~\cite{grabler_generating_2009}, and smart home control~\cite{li_programming_2017}. Some prior systems also combine PBD with other programming techniques like natural language programming~\cite{allen_plow:_2007, li_pumice:_2019} and visual programming~\cite{chasins2018rousillon}. While the implementation of the current work is developed on the \textsc{Sugilite} platform~\cite{li_sugilite:_2017}, the approach should be generalizable to other GUI-based PBD systems.

\subsection{Sharing and Cooperations in End User Development}
The sharing of PBD scripts has been discussed in prior work. Leshed et al. outlined two main motivations for sharing end-user-developed PBD task automation scripts in CoScripter~\cite{leshed_coscripter:_2008}: \textit{(i)} for tasks that are tedious and are performed frequently by many users, the main motivation is efficiency; and \textit{(ii)} for tasks that are complex or hard to remember, the shared scripts are also used as the medium for sharing \textit{how-to} knowledge among users. A later field study of CoScripter scripts~\cite{bogart_end-user_2008} revealed that users had shared CoScripter scripts for automating a wide range of work-related and non-work-related tasks, which affirmed the need for sharing end-user-developed scripts. The same study also confirmed that making scripts that were mainly intended for self-use be publicly available can lead to serendipitous reuse by others~\cite{bogart_end-user_2008}.

The privacy concern in PBD script sharing has been reported by several studies. Email and interview studies for CoScripter~\cite{leshed_coscripter:_2008} showed that the privacy concern was a main barrier to sharing. Participants reported that they decided to make scripts private because they worried about having personal information embedded in scripts. Studies and discussions for other PBD systems such as ActionShot~\cite{li_heres_2010},  RePlay~\cite{Fraser:2019:RCP:3290605.3300527}, and \textsc{Sugilite}~\cite{li_sugilite:_2017} also identified similar privacy concerns. Another prior study investigated the privacy risks of running trigger-action EUD scripts for the script users~\cite{Surbatovich2017Some}. Our work focuses on the privacy risks of sharing PBD scripts for the script authors. 

A few prior approaches have tried to address the privacy issue in sharing PBD scripts. CoScripter~\cite{leshed_coscripter:_2008} used a ``personal database'' where end user developers could manually enter their personal information such as name, phone number and email address. Whenever these entries appear in the recorded scripts during demonstrations, they are replaced with named variables. During execution, these named variables are populated with data from the script consumer's personal database. An significant drawback of this approach is that it requires users to manually create personal databases, which requires significant effort. Personal databases also only cover the most common and the most obvious types of personal information, and only handle situations where database entries show up as exact string matches in the shared scripts. Another approach is to display operations in a script as a list of textual steps (e.g., ActionShot~\cite{li_heres_2010}) or in visual programming blocks (e.g., Rousillon ~\cite{chasins2018rousillon} and Helena~\cite{chasins2017skip}) before sharing, and have users check if any personal information is included. This approach also requires manual intervention from users, and is not scalable to recent PBD systems like \cite{li_sugilite:_2017, li_programming_2017, li_pumice:_2019, sereshkeh2020vasta} that collect a large amount of contextual information along with the operations. In contrast, \textsc{Pinalite} seeks to minimize user effort, to cover a wide range of explicit and implicit personal data, and to support processing GUI contextual meta information embedded in automation scripts.

The tailoring of software artifacts is another important aspect of cooperative EUD. Some prior work in this area focused on the cooperation between end users, and sometimes professional developers or ``local experts'' during the development process (e.g., \cite{wulf_lets_1999,Gantt:1992:GGP:142750.142767,Mackay:1990:PSC:99332.99356,nardi1993small}), which we do not consider in this paper. In the scope of this paper, we assume the development is done individually. Pipek and Kahler proposed three levels of intensity of user ties in collaborative tailoring (``shared use'', ``shared context'', and ``shared tool'')~\cite{Pipek2006}. The scenarios discussed in our paper fall into the ``shared use'' level, where ``the common denominator for cooperation'' is the tool usage itself~\cite{Pipek2006}. Each shared script can automate a type of task by invoking and controlling an app (or multiple apps) that both the script author and the script consumer have. The script author and the script consumers do not share the same task instances, the same workspace, or the same context, nor is there any data exchange between them beside the script itself. \looseness=-1

\subsection{Identifying Personal \& Private Information in Data Sharing}
Our approach is also related to prior literature from the privacy and security research communities on detecting personal information leaks in shared data. For example, systems like PrivacyProxy~\cite{DBLP:journals/corr/abs-1708-06384}, Agrigento~\cite{continella2017obfuscation}, MobiPurpose~\cite{jin2018they}, and LeakDoctor~\cite{Wang:2019:LTA:3323054.3314415} detect leaks of personal information by analyzing the contents of outbound network packets. Agrigento~\cite{continella2017obfuscation} and LeakDoctor~\cite{Wang:2019:LTA:3323054.3314415} used black box differential analysis, where they test if the outbound network traffic changes when values for personal information of interest (e.g., user id, location) are changed. This approach will work with any app, but can only protect a list of pre-specified common types of personal information (e.g., location, user id), not the non-obvious ones we discussed in the Introduction (e.g., the user's reward point balance and the user's order history as shown in Figure~\ref{fig:example_leaks}). TaintDroid~\cite{enck2014taintdroid} labels data from privacy-sensitive sources (e.g., GPS, contacts) and tracks how they flow within apps and leave the system, which also only works with limited pre-specified common types of personal information. Systems like SUPOR~\cite{huang_supor_190947} detect sensitive information from user text inputs in mobile apps using natural language processing (NLP) techniques. This approach also only deals with pre-specified common types of personal information. Those methods are limited when dealing with the long-tail of uncommon types of personal information, and can be prone to re-identification attacks~\cite{el2011systematic, benitez2010evaluating}---the way to identify the user by combining multiple pieces of seemingly less-sensitive information.

\textsc{Pinalite}'s approach for identifying personal information leaks is similar to the approach used in PrivacyProxy~\cite{DBLP:journals/corr/abs-1708-06384} -- both systems identify potential personal and private information based on the uniqueness of the data. This approach does not limit the types of personal information that can be detected, providing coverage for more types of personal information. Compared with PrivacyProxy, which identifies private information in network traffic packets, \textsc{Pinalite} uses a new approach of estimating uniqueness for information shown on app GUIs (more details in Section~\ref{sec:identify}). \textsc{Pinalite} also rebuilds obfuscated information at runtime on the script consumer's side using GUI information from the script consumer's local apps (see Section~\ref{sec:executing}). 


Another way to protect personal information is to monitor when, how, and what granularity of personal information are accessed and shared by apps~\cite{Li:2017:PET:3139486.3130941}. This approach requires instrumenting each individual app, which is not scalable for our use case because we want to handle PBD scripts made with any arbitrary third-party app. In contrast, since \textsc{Pinalite} focuses on information extracted from the GUIs of apps, it does not require any special modification on target apps, and should work with most native Android apps  (with a few exceptions, such as apps with custom graphic engines like games).

\section{Background}
The implementation of \textsc{Pinalite} builds on our previously-published open-sourced \textsc{Sugilite} PBD system~\cite{li_sugilite:_2017}. This section explains the structure of \textsc{Sugilite} scripts and their corresponding metadata, and the possible types of privacy leaks in them, providing the background for the design, development, and evaluation of our approach.

\begin{figure}
	\centering
	\includegraphics[width=\columnwidth]{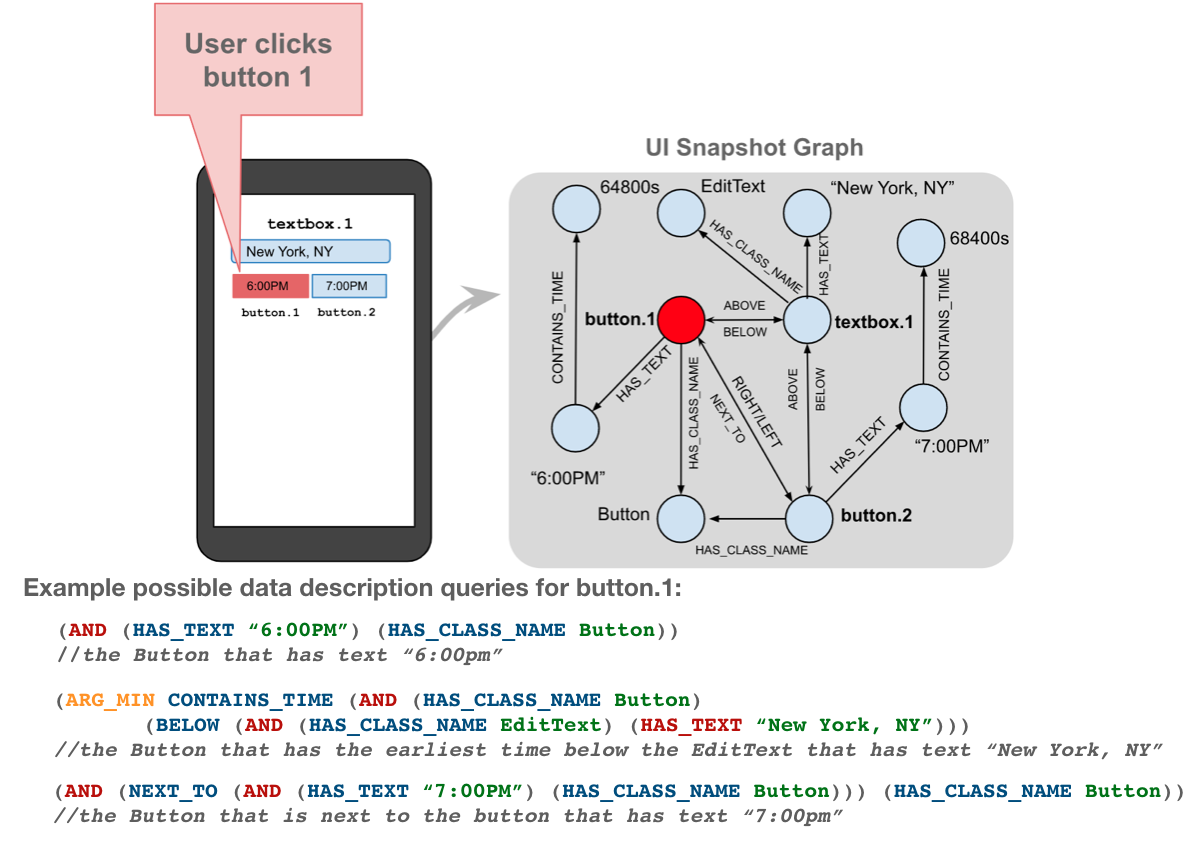}
	\vspace{-0.6cm}
	\caption{A simple example GUI, its corresponding UI snapshot graph, and some example possible data description queries for \texttt{button\_1} on the graph (highlighted in red).}
	\label{fig:ui_snapshot_graph}
	\vspace{-0.2cm}
\end{figure}

\subsection{\textsc{Sugilite} Scripts}
\textsc{Sugilite} scripts support task automation using the GUIs of Android mobile apps (e.g., to automate ordering coffee using the Starbucks app). Each script consists of a collection of operations, conditionals, and their corresponding metadata. In most cases, an operation represents an action to be performed on the GUI of an app, such as \texttt{CLICK} (click on a GUI element \textit{X}) and \texttt{SET\_TEXT} (set the value of the \texttt{text} field of a GUI element \textit{X} to \textit{Y}). For those operations, the first arguments (\textit{X} in the previous two examples) are used for identifying the target GUI elements from current screens for performing the actions (more details in Section~\ref{sec:gui_graph}). Some operations also have additional arguments (e.g., the second argument of a \texttt{SET\_TEXT} operation specifies the String content that the value of the target GUI element's \texttt{text} field should be set to). There are also a few types of special operations that do not represent actions on the GUI, such as \texttt{READ\_OUT} (read the \texttt{text} field of a GUI element out loud), \texttt{EXTRACT\_VALUE} (assign a variable using the value of the \texttt{text} field of a GUI element), and \texttt{PAUSE} (pause for \textit{X} seconds or until the user signals to continue). A script can have parameters (e.g., the \textit{size} of coffee and the \textit{type} of coffee for a coffee-ordering script). Each parameter may have a list of possible values (e.g., "tall", "grande" and "venti" for \textit{size} in the Starbucks app). The parameters and their possible values are extracted from GUIs during the demonstrations (more details in~\cite{li_sugilite:_2017, li_kite:_2018}).

Operations are linked and executed sequentially unless there is a conditional. A conditional \texttt{IF} includes one or two (with an \texttt{ELSE} clause) branches and a Boolean expression. At runtime, the execution depends on the evaluation result of the Boolean expression, which can involve constants or variables. The values of variables can be either provided before the script execution, during the script execution through extracting values from GUIs, or from the result of executing a subscript. (see ~\cite{li_pumice:_2019} for full details)
\begin{figure}
	\centering
	\includegraphics[width=\columnwidth]{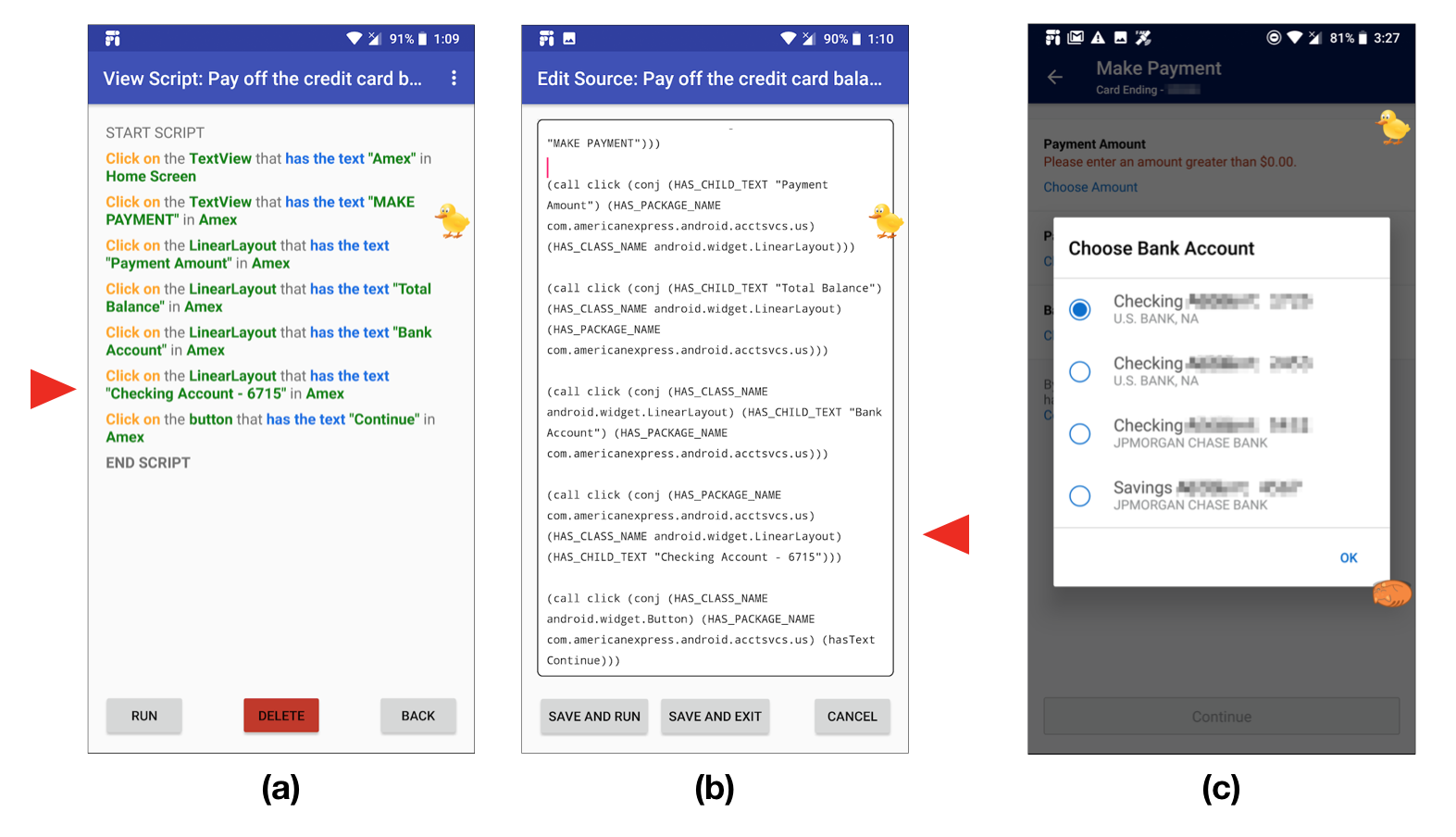}
	\caption{An example \textsc{Sugilite} script for paying off the credit card balance using the American Express app. The screenshot (a) shows the user-readable descriptions for the operations, and the screenshot (b) shows the raw source of the script. The red triangles point to data description queries that include the user's bank account information. The screenshot (c) shows the corresponding app screen for the data description query.}~
	\label{fig:screenshots1}
	\vspace{-0.5cm}
\end{figure}

\subsection{Data Descriptions and GUI Snapshot Graphs}
\label{sec:gui_graph}
In programming by demonstration, the main challenge is to infer the user's intent. From a recorded action, the system needs to determine the user's intent in performing that action so that the system can perform the desired action in a different context in the future. This process is known as creating a \textit{data description}~\cite{cypher_watch_1993,lieberman_your_2001}. For GUI-based PBD systems, data descriptions are often queries that can identify the correct target GUI element on which to perform the actions from all of the screen contents \cite{li_appinite:_2018}. 

In our case, \textsc{Sugilite} uses graphs to represent the GUI screen contents, and represents data descriptions as graph queries. Such graphs are called \textit{UI snapshot graphs} in \textsc{Sugilite}~\cite{li_appinite:_2018}. Figure~\ref{fig:ui_snapshot_graph} shows a simple example GUI, its corresponding UI snapshot graph, and a few different example queries for describing the same button in the GUI. Each graph consists of a collection of edges represented as triples denoted as $(s,p,o)$. The subject $s$ and the object $o$ are two entities, and the predicate $p$ is a directed edge from $s$ to $o$ representing a relation. Each GUI element (known as a \textit{view} in Android) is represented as an entity in the graph. They are connected to entities that represent the properties of the GUI elements, such as class types (e.g., button, checkbox, textbox)(called \texttt{HAS\_CLASS\_NAME}), on-screen bounding box coordinates (\texttt{HAS\_SCREEN\_LOCATION}), text labels (\texttt{HAS\_TEXT}), whether the elements are clickable, scrollable or focused (\texttt{IS\_CLICKABLE}, \texttt{IS\_SCROLLABLE}, and \texttt{IS\_FOCUSED}), and app-developer-assigned view IDs (\texttt{HAS\_VIEW\_ID}). The entities for GUI elements are also connected to each other based on their hierarchical relationship (i.e., \texttt{HAS\_PARENT} and \texttt{HAS\_CHILD}) and spatial relationship (e.g., \texttt{ABOVE}, \texttt{BELOW}, \texttt{LEFT}) in the current GUI layout. \textsc{Sugilite} also extracts some basic semantic relations for String entities in the graph that have easily identifiable structures (e.g., \texttt{CONTAINS\_PRICE}, \texttt{CONTAINS\_DATE}). \textsc{Sugilite}'s UI snapshot graphs are generated from the hierarchical UI trees provided by Android's Accessibility API, but this approach should also apply to other types of hierarchical mobile, desktop or web GUI representations (e.g., the DOM tree in HTML).\looseness=-1


The UI snapshot graphs at the time of demonstrations are saved and stored with their corresponding operations and data description queries in \textsc{Sugilite} scripts for a few reasons: first, they are needed to support the parameterization of scripts. In \textsc{Sugilite}, parameters and their possible values are identified using a combination of the user's natural language utterance and the app GUIs during the demonstration. For example, suppose the user selects an option with text ``venti'' from a menu during a demonstration for a task with spoken command ``order a venti cappuccino.'' \textsc{Sugilite} would identify ``venti'' as a parameter in the utterance, link it to the corresponding GUI action in the demonstration, and extract other items in the same menu (e.g., ``grande'' and ``tall'') as alternative possible values for this parameter from the UI snapshot graph. (see~\cite{li_sugilite:_2017,li_kite:_2018} for more details on parameterization).


Second, UI snapshot graphs are used to help with modifying and extending existing scripts. When the user needs to, for example, switch a data description query in use (e.g., Figure~\ref{fig:ui_snapshot_graph} shows three different data description queries that can be used to refer to the same UI element, the user may want to switch from one to another in an existing script), the original UI snapshot graph can be used to generate possible alternative data description queries, and validate the new data description query to ensure that it matches the user's intended target UI element for the action (described in detail in~\cite{li_appinite:_2018}). 

Third, UI snapshot graphs are also used for error recovery in case of execution errors. During the automation execution, when the system cannot find the target UI element for the next operation using the original data description query, the system will verify if the phone is on the correct screen of the app by comparing the current UI snapshot with the saved UI snapshot from the original demonstration. If the phone is at a different screen, it indicates a previously unseen new app state (or the app's UI has changed), so the system will prompt the user to demonstrate what to do in this new situation. For example, the Uber app will show a confirmation screen during price surges. If this new screen was not seen during the demonstration (as determined by comparing the UI snapshot graphs), the user can show how to handle this new situation by demonstration.



\subsection{Sources of Personal Information Leaks}
\label{sec:sources}
This section explains possible sources of potential information leaks in shared \textsc{Sugilite} scripts and their embedded UI snapshot graphs using examples.

\subsubsection{Data Description Queries}
\texttt{HAS\_TEXT} and \texttt{HAS\_CONTENT\_DESCRIPTION} queries used in data descriptions may contain personal information of the script author. \texttt{HAS\_TEXT} is used when the target UI element for an operation is identified using its displayed text string (the \texttt{text} field), and \texttt{HAS\_CONTENT\_DESCRIPTION} is used when the target UI element is identified using its \texttt{contentDescription} field, which is often used as the accessibility alt-text label for graphical UI elements. In many app GUIs, the contents for these fields are dynamically generated with the script author's personal information (e.g., account number, name, address---Figure~\ref{fig:example_leaks} shows a few examples). Therefore, when data description queries containing them used in the operations are subsequently shared with other users, the personal information is leaked. Figure~\ref{fig:screenshots1} shows an example data description query that leaks a script author's personal information. The data description queries include the last four digits of the user's bank account number, therefore the system needs to obfuscate personal information used in data description queries prior to sharing the script.

\subsubsection{Parameter Values}
Another possible source of personal information leaks is the list of possible values for parameters. As explained in Section~\ref{sec:gui_graph}, \textsc{Sugilite} infers the possible values for parameters from app GUIs when the script author chooses an item from a menu by extracting the alternative items that the script author could have chosen from the same menu. Therefore, personal information can be leaked when these items are personal to the script author. For example, when a script author records the ``pay off the credit card balance'' script (shown in Figure~\ref{fig:screenshots1}), an operation can be choosing the bank account. The PBD system can infer ``bank account'' as a parameter, and include all the stored accounts, which are private to the script author, as possible values for this parameter in the script. To prepare the script for sharing, the system needs to obfuscate such personal information from all parameters' list of possible values. \looseness=-1

\subsubsection{UI Snapshot Graph Texts}
As explained in Section~\ref{sec:gui_graph}, the UI snapshot graph for an app screen contains all the text strings visible on the screen. Even when the corresponding UI elements are not operated on, the text strings can still get included in the shared script to support future script modification and error handling. Although the UI snapshot graph is usually hidden from the script consumer, we do not want there to be the possibility that a script consumer could extract it. Therefore, the system also needs to examine the stored UI snapshot for each operation and obfuscate all potential personal information before sharing a script.

\subsection{Threat Model}
\label{sec:threat_model}
\tlrevision{We assume the authors of shared scripts are anonymized, and the main goal of the attackers is to obtain enough personal information about the script authors to identify them. We assume the developers of the underlying apps used for demonstrations are generally honest and well-intentioned. The local storage of \textsc{Pinalite}, the client-server communication, and the server storage are all encrypted using state-of-art techniques. The attackers are able to access to all the shared scripts, as well as upload a small number of scripts each with a small number of text fields to the server for the server-computed hash values before the server-side anomaly detection mechanism blocks them (see discussions in Sections~\ref{sec:identify}~and~\ref{sec:limitations}).}


\section{Our Approach}
In order to preserve user privacy in sharing end user developed scripts in GUI-based PBD systems, we present a new mechanism that can identify and mask potential personal and private information. To achieve this, we identified the following design goals:

\begin{enumerate}
\item The approach should require minimal user effort and intervention. In particular, it should not require script authors to manually tag personal information from the large amount of collected data. 
\item The approach should retain the transparency, readability, robustness, and generalizability of the original scripts wherever possible. 
\item The approach needs to detect and protect a wide range of common and uncommon, explicit and implicit types of personal information. Some of personal information may be previously unknown to the system, and dynamically generated by third-party apps. 
\item The approach should work with PBD scripts on general task domains on existing non-modified third-party app GUIs without the system having prior knowledge about the task domain or the underlying apps. 
\item To ensure data security, sensitive personal information should not leave the user's device (that is, the private data should stay on the phone and not be sent to the server storing scripts or, even worse, to another person's phone).
\end{enumerate}


\begin{figure}
	\centering
	\includegraphics[width=0.8\columnwidth]{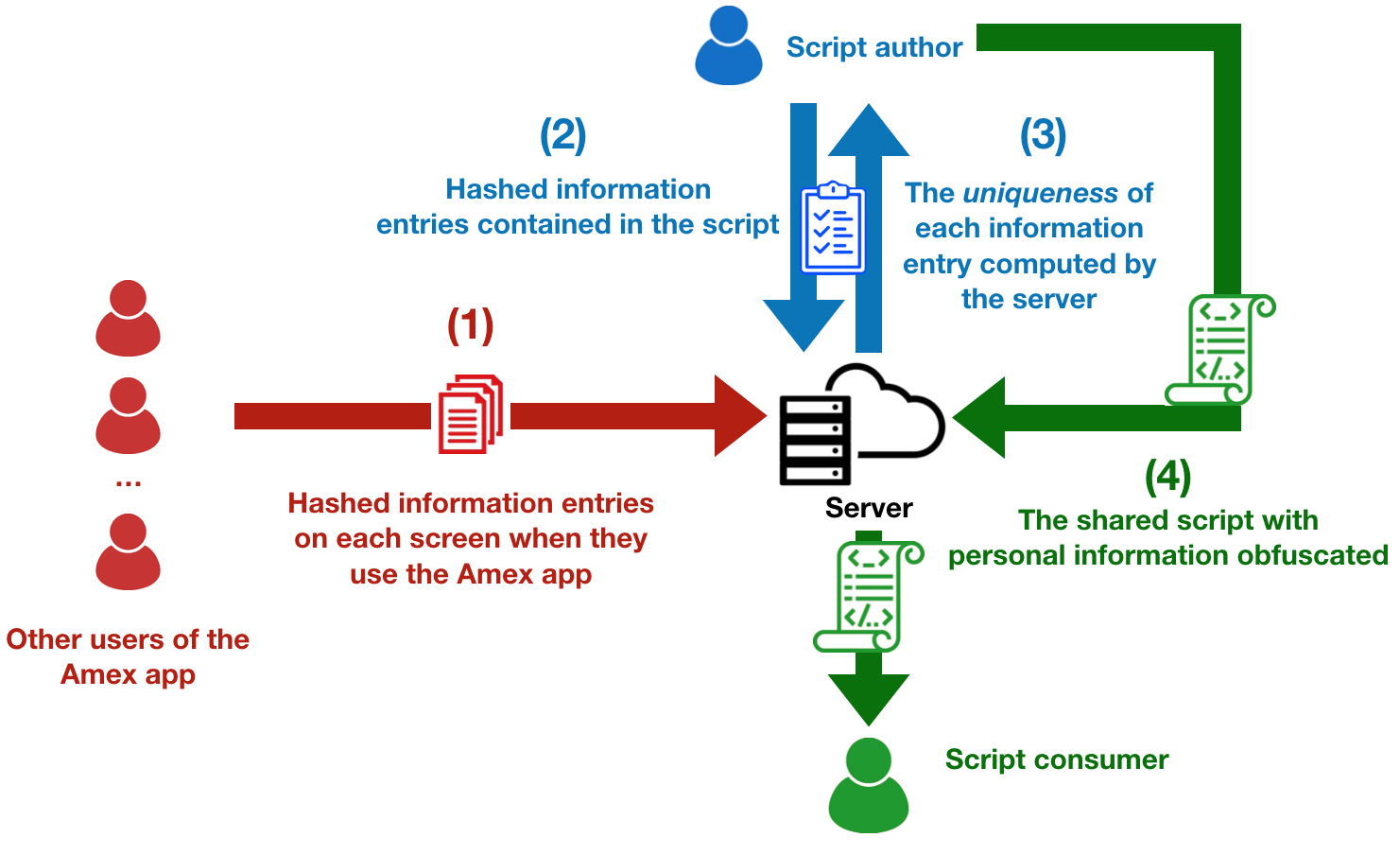}
	\caption{The diagram shows the pipeline of how our approach collects and aggregates information entries in an example third party app, uses those to compute the uniqueness for information entries before sharing a script, and obfuscates potential personal information in the shared script.}~\label{fig:architecture}
	\vspace{-0.2cm}
\end{figure}

\subsection{Overview}
Figure~\ref{fig:architecture} illustrates the pipeline of our approach. On a high level, \textsc{Pinalite} identifies potential information in a PBD script based on the \textit{uniqueness} of the information among app users. Intuitively, if a piece of information has been seen by most users, it is likely to be public, otherwise it is possibly personal. In GUI-based PBD systems, we consider an ``information entry'' to be a combination of a content (i.e., a string), and its app context (i.e., the context in the app where this string appeared). Section~\ref{sec:extracting} will discuss more about why we define an information entry this way, and how we extract them from apps.\looseness=-1

To determine the uniqueness of information entries, \textsc{Pinalite} collects and aggregates information entries in third-party mobile apps from many users in the background (Step 1 in Figure~\ref{fig:architecture}). Potential personal information on GUIs should be different for different users (e.g., everyone should have different balance, account numbers, addresses etc. in the Amex app), but other information should stay the same (e.g., the "next" button and the "contact us" number in an app). To protect user privacy, \textsc{Pinalite} only collects the \textit{hashes} of information entries, which are sufficient for determining their uniqueness. The details of this process and the definition of ``uniqueness'' used in our approach are described in Section~\ref{sec:identify}.

Before sharing a script, \textsc{Pinalite} uploads the hashes of all information entries in the script (Step 2 in Figure~\ref{fig:architecture}). The server computes and returns the uniqueness of all these information entries (Step 3 in Figure~\ref{fig:architecture}). The information entries that are more unique are considered likely personal, and therefore obfuscated using hashes by default. But the script author can also manually label and unlabel personal information through an interactive interface. Once the author confirms that all personal information is correctly obfuscated, they can go ahead and share the script (step 4 in Figure~\ref{fig:architecture}).\looseness=-1

After a different user, who we will call the script \textit{consumer}, downloads a script to their personal device, \textsc{Pinalite} supports rebuilding obfuscated fields in the script locally using the GUI contents from the corresponding target apps on the script consumer's phone. This improves the transparency, readability and robustness of shared scripts, and allows the script consumer to take advantage of the existing parameterization in the script. Section~\ref{sec:executing} will discuss this topic in more detail. \looseness=-1

\subsection{Information Entries as App Context-Content Pairs}
\label{sec:extracting}
In GUI-based PBD systems, the personal information in the scripts comes from the GUIs of the target apps. Potential personal information displayed on app GUIs can get included in the data description queries, the possible parameter values, and the UI snapshot graphs during the demonstration, as discussed in Section~\ref{sec:sources}. While some kinds of GUI-displayed information are explicitly private on their own (e.g., user phone numbers), others are only private when accompanied by the contexts of app GUIs. For example, the string ``New York'' showing up on the checkout screen of the Starbucks app can be personal and private information since it gives away the user's location, but the same string showing up on the home screen of NYC Transit app is not personal because the app would display the string ``New York'' for all of its users. Even a phone number might not be private information when it belongs to a business. For example, every user will see the same phone number when visiting the ``Contact Us'' page in the Amex app.

Therefore, in \textsc{Pinalite}, we consider an information entry as a combination of a content and its app context, and use the uniqueness of such information entries to determine whether they are likely personal information. 

In the implementation, the collected app context contains the unique identifier of an app: \textit{package name}, and the unique identifier of a screen within an app: \textit{activity name}. During the background data collection and the recording of a script, \textsc{Pinalite}'s client side software collects all text strings visible on the screen along with their app contexts as \textit{app context-content pairs}. That is, an app context-content pair denotes ``the information $X$ is displayed on the screen $Y$ in the app $Z$.''

\subsection{Background Data Collection}
\label{sec:data_collection}
After \textsc{Pinalite}'s client side software is installed on a user's phone, and given the appropriate permissions, it begins collecting information in the background, which it sends to the server (Step 1 in Figure~\ref{fig:architecture}). On each user's phone, when not recording, the client side software collects app context-content pairs (defined in Section~\ref{sec:extracting}) in the background for the current app running on the phone, and uploads the hash values of them to the server (the hashes are computed locally). Our implementation uses the SHA-512 hash algorithm for all our hash functions. It is a one-way hash function, which means that is practically infeasible to derive the original string from the hash value. The uploaded data are labelled with the user's Android \textit{Advertising ID}~\cite{google_advertising}, which is an anonymous, user-specific, unique, and resettable user ID. The server keeps a count of the number of unique users with identical information entries (i.e., how many users have seen the same information on the same screen of an app) and the number of unique users with the same app context (i.e., how many users have visited the same screen of an app). \looseness=-1

To preserve user privacy, the user only sends out the hashes of the app context and the contents in information entries, so neither any app content nor the app usage can be recovered from the uploaded data. For an information entry, the user only uploads $hash (\mathrm{app\_context}, \mathrm{content})$ to the server, so that the plain values of private information do not leave the user device. When receiving the user-uploaded hashes of information entries, the server applies another level of server-side hash with an encrypted salt before storing them (i.e., the data stored on the server for an information entry is $hash'(hash (\mathrm{app\_context}, \mathrm{content}), \mathrm{salt})$. This prevents dictionary attacks~\cite{pinkas2002securing} (i.e., the attacker enumerates the hashes of all possible content values---for example, if the attacker guesses that one of the hash values in an uploaded app UI snapshot may represent the user's city, they can enumerate the hashes for all possible city names, and compare if any matches a hash in the uploaded app UI snapshot) and rainbow table attacks~\cite{oechslin2003making} (similar to the dictionary attack, but the attacker looks up a pre-computed table of hash values for strings up to a certain length consisting of a certain set of characters), because even the attackers got access to the hashes, they would not be able to decrypt the data because they cannot compute hashes offline without the encrypted salt. 



\subsection{Identifying Personal Information for PBD Script Sharing}
\label{sec:identify}
The next step is to identify personal information in scripts. \textsc{Pinalite} uses the \textit{uniqueness} of an information entry among all users of the same to determine if it is likely to be personal. As described in Section~\ref{sec:data_collection}, the server collects and aggregates hashed information entries from app GUIs of many users. Information entries that only appear on one, or a very small number of users' devices are likely personal, while entries that appear across devices of many users are likely not. An advantage of this approach is that it makes re-identification attacks difficult. It not only protects common types of explicit personal information, but also identifies and masks any information entries that are non-common among users. As a result, information entries that are classified as non-personal are those shared by the majority of users. Therefore it would be very difficult to re-identify a user using these information entries.

Before sharing a script, \textsc{Pinalite}'s client side software for the script author first queries for the uniqueness of each information entry in the script (Step 2 in Figure~\ref{fig:architecture}). For each entry, it sends out $hash(\mathrm{app\_context}, \mathrm{content})$ in the query to the server. The server then computes the salted hash value $hash'(hash (\mathrm{app\_context}, \mathrm{content}), \mathrm{salt})$ using the encrypted salt, and counts the number of unique users with the identical information entries (i.e., the number of users who have seen this information on this screen in this app) and the number of unique users who have had the same app context (i.e., the number of users who have been to this screen in this app). 

With these counts, the server computes the uniqueness of each information entry by estimating the probability of a user has seen the same content given the user has been in the same app context. The details of the process are described below:

Let's define the observed number of unique users $F(E, U, a, c)$ as the number of users in a dataset of information entries $E$ who have had the identical app context-content pair $(a, c)$ from the user set $U$ of all the users that have contributed to $E$:

$$F(E, U, a, c)=\sum_{u \in U}[\exists e \in E : (\mathrm{user}_e = u) \land (\mathrm{app\_context_e} = a) \land (\mathrm{content_e} = c)]$$

Similarly, we define the observed number of unique users $G(E, U, a)$ as the number of users in a dataset of information entries $E$ who have had the same app context $a$ from the user set $U$ of all the users that have contributed to $E$:

$$G(E, U, a)=\sum_{u \in U}[\exists e \in E : (\mathrm{user}_e = u) \land (\mathrm{app\_context}_e = a)]$$

Therefore, for an app context-content pair $(a, c)$, the server can calculate the observed frequency $H(E, U, a, c)$ of the users who had the app context-content pair $(a, c)$ in all the users who have had the app context $a$:
$$H(E, U, a, c)=\frac{F(E, U, a, c)}{G(E, U, a)}$$

The result of $H(E, U, a, c)$ estimates the probability that a user has seen the content $c$ given that the user has been in the app context $a$. The server uses this estimated probability to indicate the uniqueness of the content $c$ in the app context $a$.

Lastly, the server tests if this probability meets a pre-specified threshold $t$ using the one-tailed exact test of goodness-of-fit~\cite{mcdonald2009handbook}. The null hypothesis is that $(a, c)$ does not meet the uniqueness threshold $t$---the probability of a user has seen the content $c$ when the user has been in the app context $a$ is greater than $t$, and the alternative hypothesis is that the probability of a user has seen the content $c$ when the user has been in the app context $a$ is less than or equal to $t$. 

If the result indicates that the probability is less than or equal to $t$ with statistical significance\footnote{We used $p<0.05$ in our implementation.}, we consider the app context-content pair $(a, c)$ to potentially contain user personal information that needs to be obfuscated. Specifying a smaller $t$ lowers the risk of false negatives (i.e., classifying personal information as non-personal) in the shared script, especially when the sample population is skewed (see Section~\ref{sec:limitations} for the detailed discussion). However, it increases the likelihood of false positives (i.e., classifying non-personal information as personal), and specifically, requires a larger number of users to collect UI snapshots from before non-personal information can be confidently identified. \looseness=-1

For example, at $t=0.5$ (which means that for an information entry to be non-personal, at least a half users who have been to this screen of this app should have seen the same information), our approach requires collected UI snapshots from 5 unique users (with the threshold of significance $p<0.05$) that \textit{all} have seen the same information before it can label this information entry as non-public. Before that, it would classify everything as potentially personal.








\begin{figure}
	\centering
	\includegraphics[width=0.8\columnwidth]{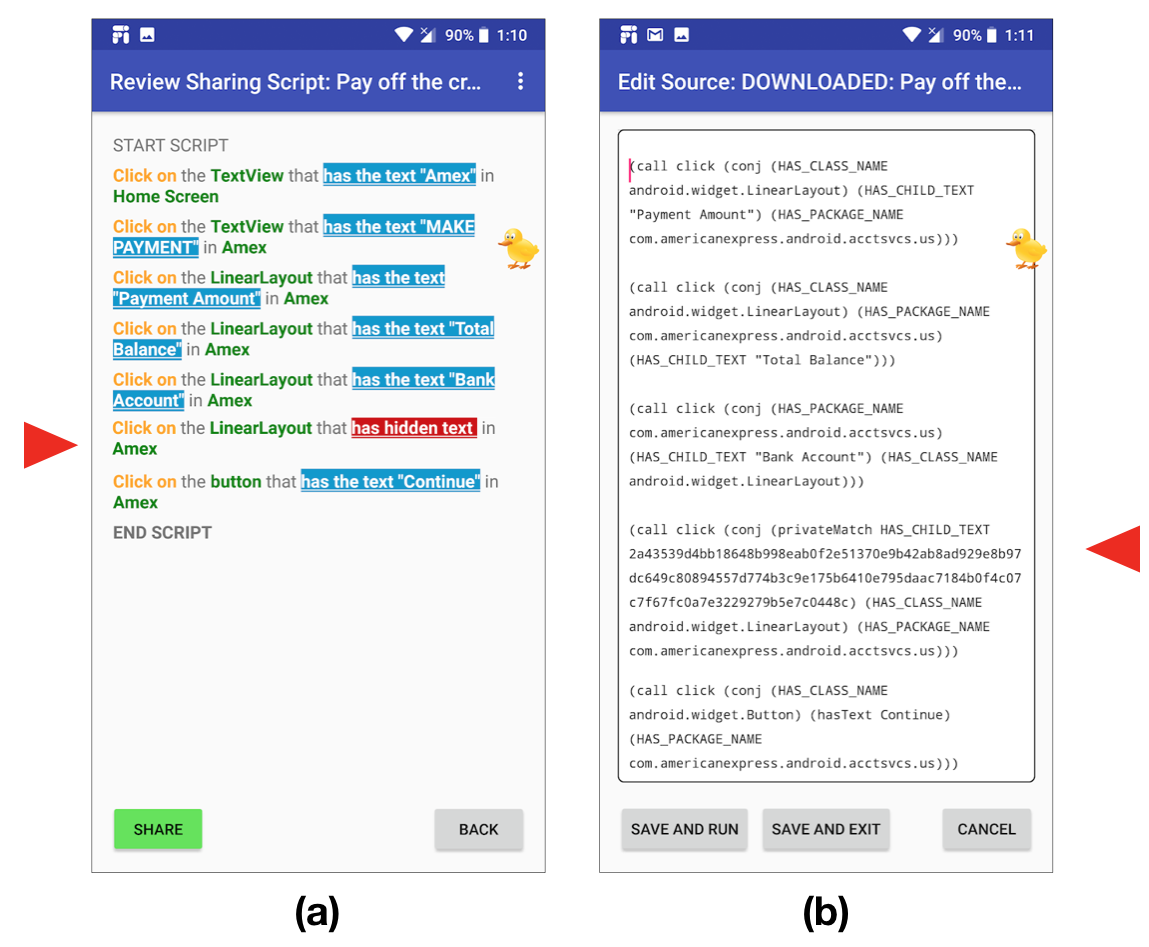}
	\caption{The result of processing the script shown in Figure~\ref{fig:screenshots1}. The left side shows the review interface for the script author, where the identified personal information is highlighted in red. The right ride shows the source of the processed script, where the content of identified personal information is replaced by its hash.}~\label{fig:screenshots2}
	\vspace{-0.2cm}
\end{figure}

\subsection{The Interactive Interface}
When the script author chooses to share a script, \textsc{Pinalite} first identifies personal information in the script using the method described above in Section~\ref{sec:identify}, and then displays the result in an interactive interface (shown in Figure~\ref{fig:screenshots2}a). For information entries that have been classified as public, the interface displays them in plain text, and highlights them in blue. Those that have been classified as potentially personal are marked as ``hidden text'' and highlighted in red. The script author can review the script to be uploaded, and manually label or unlabel private information as needed by clicking on them.  

If an information entry is labelled as public, it will be included in the shared script in plain text, otherwise the shared script will include the value of its encrypted salted hash ($hash'(hash (\mathrm{app\_context}, \\ \mathrm{content}), \mathrm{salt})$). Server-computed salted hashes are used for obfuscating the shared scripts in order to prevent dictionary attacks and rainbow table attacks, as discussed in Section~\ref{sec:data_collection}. Once the script author confirms the script, \textsc{Pinalite} will upload the obfuscated script to the server for sharing.


\subsection{Rebuilding Scripts with Obfuscated Information by Script Consumers}
\label{sec:executing}

For \textsc{Sugilite} scripts, obfuscating string values in data description queries or parameter values by hashing does \textbf{not} necessarily prevent executing scripts with hash values in data descriptions. Data description queries in \textsc{Sugilite} compare the \textit{equality} of string references in the query (e.g., click on the button that has text ``next'') with text labels of GUI elements on the screen. With hashed values, the comparison can still be done on the server side, because the hash values in the scripts are originally computed by the server using the encrypted salt. \textsc{Pinalite}'s client side software can send the hashed UI snapshot to the server during the script execution, from which the server can return which, if any, of the elements on the current UI snapshot matches the hashed query.

However, in practice the script execution using only equality checks often fails for the consumers of shared scripts when the data description queries contain the hash values of string contents that are personal to the original script author. When \textsc{Sugilite} executes an action, it will not be able to find the target UI element because the string content used in the data description query can match anything in the app on the script consumer's phone. For example, the script in Figure~\ref{fig:screenshots1} will not work with a different user, because the new user presumably does not have an account with the same account number. Therefore the system needs to rebuild the obfuscated data descriptions using the script user's local information.

There are some additional motivations for rebuilding obfuscated data descriptions: first, it helps improve script transparency. It allows script consumers to understand script behaviors, to check to see if the script fits the script consumer's needs, and allows the script consumer to modify and extend the scripts when needed. Second, as discussed earlier, comparing salted hash values can only be done on the server side. Rebuilding the obfuscated fields in scripts allows the script consumer to execute them on the consumer's phone without the help from the server. Lastly, when the script parameter values are personal to the script author, rebuilding them on the script consumer's side allows the script consumers to take advantage of the parameterization with their own personal options. For example, a script consumer can replace the data description query in the highlight operation in the screen shown in Figure~\ref{fig:screenshots1} with their own account numbers, and use them as possible parameter values for the parameter ``account number'' so that this script can be used for all of the script consumer's bank accounts.


\subsubsection{Rebuilding Obfuscated Fields}
Here we explain how \textsc{Pinalite} rebuilds the obfuscated fields in shared scripts locally on the consumer's phone. In the earlier example shown in Figure~\ref{fig:screenshots1}, when a script author records a script that pays off the credit card balance using the Amex app, one step is to choose an account to pay from. This action is recorded as ``Click on the LinearLayout that has the text `Checking Account ...1234 in Amex''. Since the author chooses this account from a menu that also contains options for the author's other accounts, the PBD system can recognize ``account'' as a parameter, and save the other account options as possible values for this parameter so that this script can later be used for paying off the credit card balance using the other accounts (see more details about the script parameterization process in~\cite{li_sugilite:_2017, li_kite:_2018}).  However, when \textsc{Pinalite} prepares the script for sharing, both the original data description text (``Checking Account ...1234'') and the alternative options will be obfuscated because these information entries are all personal to the script author. When script consumers download this script, all they can see is ``Click on the LinearLayout that \textit{has hidden text} in Amex'' (shown in Figure~\ref{fig:screenshots2}a). Under the hood, the hidden text represents a salted hash of the original text in the source, as shown in Figure~\ref{fig:screenshots2}b. This script will not execute correctly, because the script consumer would not have an account whose text label matches the hash of the original account text label of the script author. The parameter mechanism would be broken too for the same reason.

A prerequisite to our solution is to add a step in the script recording mechanism on the script author's side. Each time the script author demonstrates an operation, in addition to generating the main data description that best reflects the script author's intention, the recording mechanism \textit{also} extracts an alternative data description query that uses no potential personal information on app GUIs. As explained in Section~\ref{sec:gui_graph}, each app screen is represented as a GUI snapshot graph. Therefore, the alternative data description needs to be a graph query that (1) uniquely identifies the target UI element in the graph, and (2) does not use the (\texttt{HAS\_TEXT}) relation on strings containing any potential personal information. For the credit card payment example used above, an alternative to the original data description ``Click on the TextView that has the text `Checking Account (...1234) in Amex'' can be ``Click on the first TextView in the ListView below the TextView that has the text ``Choose Bank account'' in Amex.'' In this alternative query, the text ``Choose Bank account'' is not personal, because everyone who has visited the account selection screen in the Amex app can see the same text on the screen. \looseness=-1

On the script consumer's phone, \textsc{Pinalite} rebuilds the script at the runtime using the script consumer's local app GUI information. When executing a script, when \textsc{Sugilite} reaches the account selection menu, it cannot find any screen element with text that matches the hash value in the script. At this time, the system uses the previously generated alternative data description query to locate the account selection menu on the screen, replaces the ``hidden text'' data description with the new value (e.g., ``Checking Account (...2345)'' found locally on the consumer's phone from the matched UI element, and extracts other menu options from the GUI (e.g., ``Saving Account (...3456)'') to be used as the new possible parameter values \textit{for this specific script consumer}. This allows the script consumer to see the data description queries revealed in the script (e.g., the script consumer's own account information for the script in Figure~\ref{fig:screenshots2} instead of ``hidden text''), and to take advantage of the parameterization in the script using their own personal information.

An underlying assumption of this approach is that the \textsc{Sugilite} operates on the script consumer's personal device, and the script consumer does not share the device with others. Our approach assumes that the script consumer \textit{should} have access to any information the system sees in app GUIs locally on their phone, and therefore the information can be safely used to rebuild the script consumer's personal copy of the automation script. Similarly, other people should not have access to the script consumer's personal copy of the script stored locally on the phone, since it may contain the script consumer's personal information.





\section{Evaluation}
We evaluated our approach on three research questions: 

\begin{enumerate}[label={RQ\arabic* },wide = 0pt, font =\textbf]
    \item How accurate is \textsc{Pinalite}'s approach in identifying personal information in GUI-based PBD scripts?
    \item How well can script consumers understand the shared scripts after they are obfuscated using \textsc{Pinalite}?
    \item How comfortable do script authors feel about sharing their scripts with \textsc{Pinalite}?
\end{enumerate}

We investigated the first question through an accuracy evaluation, and the last two questions through a lab user study.

\subsection{Accuracy of Identifying Potential Personal Information}
\subsubsection{Method}

We selected 15 popular Android apps from various task domains in the Google Play Store. For each app, we came up a task that (1) reflects the main purpose of the app, (2) is reasonable to be automated using an agent, and (3) interacts with some personal information of the user. We performed each task with 5 different ``simulated users'', and collected the corresponding information entries from these UI snapshots. For each simulated user, We created a new app account using different a profile, and spoofed a new GPS location. (Note that we did not select, for example, banking apps and medical apps in this evaluation because it was difficult to create new accounts for simulated users in these apps.)  We then created a \textsc{Sugilite} script for each target task by demonstration, and used \textsc{Pinalite} to identify personal information in the script with the uniqueness threshold $t=0.5$. 

For each script, we extracted all of the information entries contained in its data description queries, parameter values, and UI snapshots. We evaluated the performance of our approach on these scripts in terms of \textit{precision} (the fraction of information entries that contain personal information among the ones classified as ``potentially personal''), \textit{recall} (the fraction of information entries that contain personal information that were correctly classified as ``potentially personal''), and \textit{accuracy} (the fraction of information entries that were correctly classified) of identifying personal information by comparing the result against expert human judgements. We used two experts to independently label whether each information entry contains any personal information. The two judges reached \textit{substantial agreement}~\cite{landis1977measurement}, with the percent $\mathrm{of\ agreement}=87.8\%$ and $\mathrm{Cohen's}\ \kappa=0.63$.

\begin{table}[thb]
	\begin{small}
		\begin{center}
			\begin{tabular}{|l|l|l|l|l|l|l|}
				\hline
				App & Task & $n$ & $n_{\mathrm{personal}}$ & Recall & Prec. & Accu. \\
				\hline
			    Starbucks & Order a cup of coffee for pick up & 58 & 11 & 1 & 0.38 & 0.83 \\
			    Uber & Request a ride & 128 & 35 & 1 & 0.74 & 0.69 \\
			    Marriott & Book a hotel room & 140 & 8 & 1 & 0.44 &  0.93 \\
			    Papa John's & Order a pizza for delivery & 142 & 5 & 1 & 0.19 & 0.85 \\
			    AccuWeather & Check the current weather & 59 & 3 & 1 & 0.17 & 0.75 \\
			    OpenTable & Make a restaurant reservation & 116 & 12 & 1 & 0.26 & 0.71 \\
			    Kayak & Book a flight & 89 & 21 & 1 & 0.51 & 0.78 \\
			    AMC Theatres & Buy a movie ticket & 91 & 6 & 1 & 0.12 & 0.53 \\
			    Zillow & Check the homes for sale in the area & 42 & 4 & 1 & 0.25 & 0.71 \\
			    Lyft & Request a ride & 68 & 37 & 1 & 0.86 & 0.85 \\
			    Cinemark Theatres & Buy a movie ticket & 193 & 9 & 1 & 0.11 & 0.60 \\
			    Chipotle & Order food for pickup & 136 & 34 & 1 & 0.89 & 0.94 \\
			    Yahoo! Sports & Check sports scores & 86 & 13 & 1 & 0.34 & 0.71 \\
			    Dunkin' & Order a drink for pick up & 76 & 19 & 1 & 0.83 & 0.95 \\
			    Uber Eats & Order food for delivery & 63 & 13 & 1 & 0.41 & 0.73 \\
				\hline
			\end{tabular}
		\end{center}
	\end{small}
	\caption{The result of the evaluation, showing the app name, the task, the number of information entries collected ($n$), the number of information entries that contained personal information ($n_{\mathrm{personal}}$), the recall, the precision, and the accuracy for each script.}
    \vspace{-0.6cm}
	\label{tab:result1}
\end{table}

\subsubsection{Results and Discussion}
\label{sec:study1_discussion}
Table~\ref{tab:result1} reports the result of the evaluation.  First of all, note the large number of strings in each script (column $n$), ranging from 42 to 193 strings---too many for script authors to comfortably review manually. All 15 scripts included some personal information of the script author (column $n_{\mathrm{personal}}$). Our approach achieved a perfect recall---that is, all personal information entries were indeed correctly identified as potentially personal, and therefore obfuscated. No personal information was leaked in the processed scripts in our evaluation. This result suggests that our approach is adequate in preserving the script author's personal information in end-user developed PBD scripts. 

On the other hand, the precision of our approach was lower in some task domains. That is, our approach sometimes classifies non-personal information as personal information. This trade-off was a deliberate design decision we made---our approach seeks to minimize false negatives by not limiting to pre-specified common types of personal information such as phone numbers, email addresses, and birthdays. Instead, it by default obfuscates all ``unique'' information, defined as information that is not seen by most people who have visited the same screen in the same app. This approach ensures a higher degree of privacy for users. A trade-off of this approach is that it may introduce false positives by obfuscating information that people might not consider private. When we looked into the false positives in our evaluation, they were mostly (1) personalized recommendations and advertisements---fields that show different contents for different users such as movie recommendations in AMC and Cinemark apps; and (2) dynamic information that depends on the time and location, such as the temperature readings in AccuWeather, the available timeslots for restaurants in OpenTable, the date selections in Marriott, and the latest sports news in Yahoo! Sports. It is also worth noting that these kinds of false positives are unlikely to impact the function and the usability of the shared PBD scripts. These fields are almost never used in data descriptions for script actions, and are rarely useful for script generalization and error handling because they are not consistent across multiple instances of script execution. \looseness=-1

Our approach successfully identified many kinds of uncommon and seemingly innocent personal information that may be combined for re-identification attacks, such as the address of and the distance to nearby locations in Starbucks, Papa John's, and AMC Theatres, the addresses of nearby real estate properties in Zillow, the amount of reward points in Starbucks and Marriott, the names of local sports teams in Yahoo! Sports, and the past order history of the script author in Uber Eats. This was in addition to the common types of explicit personal information identified in the scripts such as the the script author's name in greetings (``Good morning, Bob''), the stored script author's home address, and the script author's account name. When labeling the ground truth for personal information, we did \textbf{not} count information entries that may only reveal the script author's personal information with some prior domain knowledge, such as the list of a product offering that is only available to users in a certain region and a menu option that only shows up for users with certain reward status. In our metric, those were counted as false positives, but they were protected using our approach nevertheless.

\subsection{User Study}
\subsubsection{Participants}
We recruited 16 participants (5 women, 11 men, ages 21--32) for our user study. The user study session for each user lasted around 30-40 minutes. We compensated the participants \$15 for their time. Most (75\%) participants were students at local universities. The others worked different technical, administrative, or managerial jobs. Among the participants, there were 4 (25\%) non-programmers (i.e., those who have done none or only very light programming such as writing Excel functions), 5 (31.3\%) novice programmers (i.e., those who have taken 1-2 college level computer science classes, or with equivalent expertise, but have little ``real-world'' programming experience), and 7 (44\%) more experienced programmers. 5 (31.3\%) participants had some prior experiences with PBD systems (e.g., macro recorders in MS Office and Adobe software), while the other 11 participants (68.7\%) had no prior experience with PBD. 

\subsubsection{Methods}
For each study session, after obtaining the proper consent and having the participant fill out a demographic survey, we gave the participant a short tutorial on the general concept of GUI-based PBD, and how \textsc{Sugilite} works using its published tutorial video\footnote{\url{https://www.youtube.com/watch?v=KMx7Ea6W6AQ}}. After the tutorial, we investigated RQ2: How well can script consumers understand the obfuscated scripts?

We gave each participant 3 tasks in random order. The goal of the tasks was to simulate the experience of understanding a shared script for a script consumer. For each task, the participants saw an obfuscated script like in Figure~\ref{fig:screenshots2}a.  We used the Amex example shown in  Figure~\ref{fig:screenshots2}, and the Uber and Dunkin' scripts that were also used in the accuracy evaluation (Table~\ref{tab:result1}). The participants saw screens of running the script on the phone of a simulated user different from the script author (so that the screens would show different contents in the personal information fields). \tlrevision{The user experience in this process should be identical to that when a script user downloads and executes a shared script using \textsc{Pinalite}. We used a simulated user profile instead of having the participants create new user profiles in the study in order to avoid accidentally collecting any actual personal information from the participants.} We then asked the participant to tell us what the purpose of the script was, and what the ``hidden text'' represented in the script. Finally, each participant rated if it was easy for them to understand what each script does. 

Following the tasks, we moved to answering RQ3 (How comfortable do script authors feel about sharing their scripts with \textsc{Pinalite}?) using a questionnaire. We first described the concept of uniqueness in our approach, explained the guarantee that our approach provides (for an information entry to be non-personal, at least a half users who have been to this screen of this app should have seen the same information), and showed the review interface using the Amex script example. \tlrevision{We did \textit{not} have the participants demonstrate new scripts because (1) we wanted to exclude the factor of the usability of \textsc{Sugilite}'s recording mechanism (which has been evaluated in our previous paper~\cite{li_sugilite:_2017}) from this study; and (2) we wanted to avoid accidentally collecting any actual personal information from the participants.} Each participant then filled out a questionnaire on how much they trusted the system as a script author, how comfortable they were with sharing scripts in our system, and whether the review interface was clear about what information is hidden and what information is revealed in shared scripts. Finally, we ended the session with an informal interview on their thoughts about our approach. We specifically asked if they had any privacy concerns with our approach, and whether they could think of any scenarios where they would not be comfortable sharing scripts processed with our approach. \looseness=-1

\subsubsection{Results and Discussion}
The results for RQ2 are encouraging. Our participants found the obfuscated scripts easy to understand. After completing the 3 tasks of understanding obfuscated shared scripts, they on average rated 4.69 out of 5 (standard error of the mean $\sigma_{\bar{x}}=\pm0.12$) on a 5-point Likert scale for the statement ``\textit{I find it easy to understand what each script does}'', and on average rated 4.44 out of 5 ($\sigma_{\bar{x}}=\pm0.19$) for the statement ``\textit{The script representation is useful for me to determine if the script serves my needs}'' in the questionnaire. 

In the interview, the participants confirmed our hypothesis that script consumers would like to review the scripts from others before executing them. For example, P9 agreed they ``were concerned about using others’ scripts, [and they wanted to check] whether that contains the virus or other unwanted hidden procedures.'' However, several participants, although finding the current representation clear and easy-to-understand, also thought it to be too difficult to go through the entire script. P15 said, ``\textit{...there must be some easier way to do it.}''   P3 and P12 also worried that some words used in the script representation (e.g., ``TextView'' and ``LinearLayout'') were too technical for end users (both P3 and P12 were experienced programmers). These findings identified the research opportunity of designing an easier-to-use script representation of PBD scripts specifically for script consumers. But overall, participants did not have much of a problem with the transparency of the scripts with obfuscated fields. They found it easy to understand the purpose of each script and what each ``hidden text'' might represent as a script consumer, with the the help of referencing the GUIs of the underlying third-party apps. 

The results for RQ3 are also generally positive. 14 (87.5\%) participants reported that they felt comfortable sharing their personal scripts with others using our approach in the study. 14 (87.5\%) participants found the script review interface (shown in Figure~\ref{fig:screenshots2}a) clear about what information was private and what information was public in the scripts. However, there were 2 participants who reported that they could not trust \textsc{Pinalite} to protect their personal information. 

Some of the distrust originated from the sensitivity of certain task domains. For example, in the interview, P1 reported, ``\textit{I might not let a virtual agent do the task related to my bank account.}'' but was comfortable sharing scripts in other task domains with \textsc{Pinalite}. Another participant who reported distrust in our system, P6, said, ``\textit{It is not about this mechanism or app, I just don’t trust technology in general},'' and later added, ``\textit{but sharing with friends is acceptable.}" A few other participants, while comfortable with sharing scripts with the public using \textsc{Pinalite}, also mentioned that scripts shared only with friends can have different standards for privacy. P4 and P7 wanted to include more personalizations in their scripts to make them more effective for their friends to use. P4 said, ``\textit{I want to share with my friends about ...(some customization of coffee) since I think my recipe is delicious. But that information would be definitely hidden with your mechanism.}'' Lastly, a few participants reported non-privacy-related concerns about script sharing. P4 was not sure if his scripts are effective enough for other people to use, and P13 said ``\textit{... I don’t want to be responsible for the performance of my scripts.}” 

\section{Limitations}
\label{sec:limitations}
Our current approach has some limitations. First, the current version of \textsc{Pinalite} only processes textual information. It can not identify and obfuscate personal information from visual information in images, or other multimedia forms. This limitation does not pose a problem when working with the current version of the underlying \textsc{Sugilite} PBD system~\cite{li_sugilite:_2017}--\textsc{Sugilite}, like many other GUI-based PBD systems such as CoScripter~\cite{leshed_coscripter:_2008}, PLOW~\cite{allen_plow:_2007}, and ActionShot~\cite{li_heres_2010} do not use information about the contents of images and other multimedia forms in scripts. So they do not collect any visual information in their scripts. But for our approach to work with vision-based PBD systems (e.g., Sikuli~\cite{yeh_sikuli:_2009} and HILC~\cite{Intharah:2019:HDP:3320251.3234508}), it would need to be able to identify and obfuscate visual information as well. 

\tlrevision{Second, \textsc{Pinalite} also does not consider situations where the layout of an app encodes personal information. For example, suppose a layout structure in an app only shows up for users in a specific user group (e.g., the entire ``loan'' menu in a banking app only shows up for users with outstanding loans), then the presence of references to this GUI layout structure in the alternative queries used for rebuilding scripts may reveal personal information about the script author.}   \looseness=-1

Third, when testing whether the uniqueness of an information entry meets the threshold $t$, we assume that the population of sample users is independent. This assumption can be a problem when the sample population is \textit{very significantly} skewed. Let's look back at the example in Section~\ref{sec:extracting}, the string ``New York'' on the checkout screen of the Starbucks app should be identified as potential personal information, as it gives away the script author's current city. However, if the system encounters a skewed sample population where the vast majority of users are actually from New York, the system may classify it as a non-personal information entry, and therefore reveal it by default in future shared scripts. One way to mitigate this problem is to impose more strict thresholds for $t$ and for the $p$ value used in the statistical test. But this comes with the trade-off of having more false negatives (non-personal information classified as personal information). It is worth noting that this limitation only affects ``less differentiating'' personal information such as the current city, gender, and country when the sample population is significantly skewed. More sensitive information such as names, exact addresses, account numbers will not be impacted because these are more likely to be unique among the users. \looseness=-1

Fourth, \textsc{Pinalite} uses a server-side salted hash approach (see Section~\ref{sec:identify}) to prevent dictionary attacks. The attackers cannot decrypt obfuscated personal information by comparing the hashes appeared in scripts against offline-computed hashes for a large number of possible values because computing hashes requires the access to the encrypted salt. The server can control the access to the salted hash function and block abnormal requests (e.g., those with abnormally large numbers of strings, or from clients with too many requests). However, for fields with only a few possible values (e.g., gender), it is still possible for an attacker to enumerate all possible values in the server requests without being detected and blocked. But this problem will not affect more sensitive personal information fields like names, exact addresses, and account numbers, which have larger numbers of possible values. \tlrevision{In our threat model (Section~\ref{sec:threat_model}), we assume that the authors of shared scripts are anonymized, and the main goal of the attackers is to obtain enough personal information about the script authors to identify them. Therefore, fields with only a few possible values are not very useful (e.g., knowing a script author is a female Starbucks app user between 25-39 is not sufficient for identifying her). Since shared scripts are anonymized, and hashes are computed with app contexts (i.e., the same string will result in different hash values when appearing in different apps, or different screens within the same app), the attackers are not able to link information across multiple scripts to further deanonymize the script author either. However, when the identity of the script author is known, such as when a script author directly shares a \textsc{Pinalite}-obfuscated script with others, potential sensitive information with a small number of possible values (e.g., gender and age group) may get compromised when the attackers try enumerating the possible options with the \textsc{Pinalite} server to obtain their hashes.} 

In practice, \textsc{Pinalite} can also be combined with other methods to support identifying potential personal information at a finer granularity. For example, one can use the ``personal database'' method used in CoScripter~\cite{leshed_coscripter:_2008} to manually blacklist specific fields known to contain personal information (e.g., gender), and use regular expressions to explicitly identify personal information with fixed structures like email addresses and phone numbers. This would allow the system to obfuscate these fields using named types like ``hidden email address'' and ``hidden user name'' instead of the generic ``hidden information'', further improving the readability and generalizability of shared scripts. \looseness=-1



\section{Conclusion and Future Work}
This paper presents a new mechanism, called \textsc{Pinalite}, for identifying and obfuscating possible personal information in GUI-based PBD scripts based on the uniqueness of information entries in app GUIs. By collecting hashed samples of GUI screens from many users, \textsc{Pinalite} is able to determine if a piece of information is likely personal by estimating the probability that a user has seen this information when that the user has visited this screen in the app. \textsc{Pinalite}'s approach can also replace the obfuscated fields in a shared script locally using the script consumer's app GUI contents, preserving the transparency, readability, robustness and generalizability of the original script. \looseness=-1

The current version of \textsc{Pinalite} addresses the privacy-preserving sharing of individual scripts. We are currently exploring aggregating scripts for similar tasks from multiple users. This would allow us to collect data about task automation procedures on different phones in various environments, and create more robust scripts that can handle differences in target app versions, screen layouts, phone usage contexts and task parameters. Prior to the introduction of an effective privacy-preserving mechanism, we cannot collect runtime data of GUI-based PBD systems which contain actual script contents in the background, since the contents may contain the script author's personal information. With \textsc{Pinalite}, it may be feasible to extract runtime data of GUI-based PBD systems in the background, which would allow the PBD system to collect more data for aggregation purposes.   

We also plan to conduct a larger-scale field deployment of \textsc{Pinalite}. Through that study, we can test whether our approach in fact promotes script sharing behaviors in real usage. We will also study how script authors actually use \textsc{Pinalite} to obfuscate personal information in scripts for the purpose of sharing. 

\tlrevision{Another interesting future research direction is to better communicate with the script authors and the script users about the data they may share and the privacy expectations associated with sharing these data. This should help users become more comfortable with participating in the background data collection and sharing their scripts, while making informed decisions on whether to share a script and whether to mask/unmask a field in a shared script at the script review stage. We hope to investigate this topic as a part of our planned field deployment.}

\begin{acks}
This research was supported in part by Verizon and Oath through the InMind project, a J.P. Morgan Faculty Research Award, and NSF grant IIS-1814472. Any opinions, findings and conclusions or recommendations expressed in this material are those of the authors and do not necessarily reflect the views of the sponsors. We would like to thank our study participants and our anonymous reviewers. We also thank Haojian Jin, Michael Liu, Fanglin Chen, Sujeath Pareddy, and William Timkey for helpful discussions and insightful feedback.
\end{acks}

\bibliographystyle{ACM-Reference-Format}
\bibliography{references}

\end{document}